\documentclass[12pt,preprint]{aastex}

\shorttitle{Recent Evolution of Clusters}
\shortauthors{Plionis}

\begin{document}

\title{Recent Dynamical Evolution of Galaxy Clusters}


\author{M. Plionis}
\affil{Instituto Nacional de Astrofisica, Optica y Electronica (INAOE),
 Apartado Postal 51 y 216, 72000, Puebla, Pue., Mexico \\
and \\
Institute of Astronomy \& Astrophysics, National Observatory of
Athens, I.Metaxa \& B.Pavlou, P.Penteli 152 36, Athens, Greece}


\begin{abstract}
Evidence is presented for a recent evolution of the 
relaxation processes in clusters of galaxies, using large optical and X-ray
cluster samples. The criteria of the cluster relaxation used are
the cluster ellipticity, the ICM temperature and X-ray cluster luminosity. 
We find evidence of varying strength and significance of
all three indicators evolving with redshift for $z\lesssim 0.15$.
This result supports the view that clusters have
mostly stopped undergoing mergers and accreting matter, as expected in
a low-$\Omega_{m}$ Universe, and are now in the process of
gravitational relaxation, which reduces their flattening, their
ICM temperature (shock heated during the merging phase), and their
X-ray luminosity. These results support similar recent
claims of Melott, Chambers and Miller.
\end{abstract}

\keywords{galaxies: clusters: general - galaxies: evolution
-large-scale structure of universe}

\section{Introduction}
The present dynamical state of clusters of galaxies should contain
interesting cosmological information since
the rate of growth of perturbations is different
in universes with different matter content \citep{Pee80};
in an $\Omega_{m}=1$ universe the perturbations grow
proportionally to the scale factor [$\delta \propto (1+z)^{-1}$],
while in the extreme case of an empty universe, they
do not grow at all [$\delta =$ constant]. It is also known
that $\Omega_{m}<1$ universes behave dynamically as an 
$\Omega=1$ universe at large redshifts, while at some redshift
$z\simeq \Omega_m^{-1}-2$ perturbations stop evolving, allowing 
clusters to relax up to the present epoch much more than in an
$\Omega_{m}=1$ model, in which clusters are still forming.

Note that a flat $\Omega_{\Lambda}> 0$ model behaves as an 
$\Omega_{m}=1$ model up to a lower redshift than the 
corresponding open model, while for redshifts $z\lesssim 1$ it
behaves like an open model \citep{Lah91}. 
This implies that clusters even in this model should be dynamically 
older than in an EdS model \citep{Ri,Ev,Lac,Bei,Suwa}.

Therefore one should be able to put constraints on $\Omega$ 
from the evolution of the cluster dynamical state. One such indicator, the
cluster ellipticity, was recently proposed in \citet{MelC}; 
the more elliptical a cluster the younger it is,
since ellipticity should be correlated with the anisotropic accretion of
matter and merging \citep{Buote, KBPG}.
\citet{MelC} presented evidence for a
recent evolution ($z\lesssim 0.1$) of cluster ellipticity, defined in both
optical and X-ray bands, based on various cluster samples,
containing in total $\sim 160$ rich Abell clusters.
An earlier study, using clusters identified in the Lick map 
\citep{PBF91}, had also found
that cluster ellipticity decreases with redshift for $z\lesssim 0.1$, 
due however to
possible systematic effects involved in the construction of the data,
the authors did not attach much weight to this result.

This interesting finding was interpreted in \cite{MelC} as an
indication of a low-$\Omega_{m}$ universe, 
because in such a universe and under the hierarchical clustering scenario,
one expects that merging and anisotropic accretion of matter along
filaments \citep{Shan} will have stopped long ago. Thus the clusters should be
relatively isolated and gravitational relaxation will tend to isotropize the
clusters reducing their ellipticity, more so in the recent
times. A variety of indications do exist supporting the formation of
clusters by hierarchical aggregation of smaller units along filamentary
large-scale structures \citep{West, PB02}.

Although the cluster ellipticity is a relatively well defined
quantity, systematic effects due to projections in the optical, that 
could increase the ellipticity of spherical clusters or decrease 
the ellipticity of flat ones, or the strong central concentration of the
X-ray emitting gas (since $L_{x} \propto n^{2}_{e}$) in
the X-ray band, which in conjuction with resolution effects
could reduce the ellipticity of distant clusters, 
should be taken into account. One should
therefore use large and well constructed samples of galaxy clusters to
verify such an ellipticity trend.
Furthermore, other indicators of recent cluster evolution should
be used as well. For example, if the
ellipticity decrease with $z$ is indeed due to recent dynamical
relaxation processes, then one should expect an evolution of the
temperature of the X-ray emitting gas as well as the X-ray cluster
luminosity which should follow the same
trend as the cluster ellipticity, decreasing at recent times, since
the violent merging events, at relatively higher redshifts, will 
compress and shock heat the diffuse ICM gas \citep{Ritchi}.
Another possible indicator could be the 
cluster velocity dispersion, which naively one would expect
to increase at lower redshifts, 
since virialization will tend to increase the cluster
`thermal' velocity dispersion. Although there is a quite well defined
$L_x-\sigma$ relationship \citep{Quin, WhiteF}, unrelaxed clusters can also
show up as having a high velocity dispersion due to either possible large
peculiar velocities of the different sub-clumps \citep{Rose}
or due to the possible sub-clump virialized nature. 
Therefore, a better physical understanding of the
merging history of clusters is necessary in order to be able
to utilize the velocity dispersion measure as an evolution criterion.

In this letter, we will present further evidence for a recent
dynamical evolution of clusters, using the 
optical APM and the X-ray XBAC and BCS cluster samples and three
indicators of cluster dynamical evolution,
namely ellipticity, the X-ray cluster temperature and X-ray
luminosity.

\section{Data}
Details of the APM cluster sample and its construction can
be found in \citet{Dalton}. Here we only remind the reader that
the APM cluster catalogue is based on the APM galaxy survey, which
 covers an area of 4300 square degrees in the southern
sky and contains about 2.5 million galaxies brighter
than a magnitude limit of $b_{J}=20.5$ \citep{Mad90}.
\citet{Dalton} applied an objective cluster finding algorithm to the
APM galaxy data using a search radius of  $0.75 \; h^{-1}$ Mpc, in order
to minimize projection effects, and so produced a list 
of 957 clusters with $z \lesssim 0.15$. 
Out of these 309 ($\sim 32\%$) are ACO clusters, while 407 
($\sim 45\%$) have measured redshifts.
The APM clusters that are not in the ACO list are relatively poorer
systems than the Abell clusters, as can be verified comparing their mean 
APM richness's, determined in \citet{Dalton}.

The APM cluster shape parameters (ellipticity and position angles) were
estimated in \cite{BPM} and were used to measure the intrinsic 
distribution of clusters shapes (found to be prolate-like). 
The shape parameters
were estimated using the moments of inertia method on the Gaussianly
smoothed galaxy distribution, the angular smoothing scale of which depends
on the distance of each cluster. An optimum grid size was
selected based on an extensive Monte-Carlo procedure.
In order not to bias the ellipticity measurements
no circular aperture was used, but rather all cells above a selected
density threshold were feed into the inertia tensor. Three such
thresholds were used to provide the shape parameters at three
different scales and to test their robustness as a function of
distance from the cluster center [see details in \citet{BPM} and
\citet{KBPG}].
The shape parameters for all APM clusters as
well as indicators of the dynamical state of each APM cluster will be
published shortly (Plionis \& Basilakos in preparation).

Here we will use the ellipticities of 903 APM clusters that have
measured shape parameters (54 clusters are found in the vicinity of 
plate-holes or crowded regions and thus are excluded). 
We will also use the XBAC \citep{Ebel} and BCS \citep{Ebel1} 
ROSAT X-ray cluster samples
to test whether there is any indications of a recent evolution of the
ICM temperature. Details of the sample construction can be found in
the original papers. We only note that
the XBACs has a flux-limit of $f_x>5 \times 10^{-12}$
sec$^{-1}$ cm$^{-2}$ and contains 283 Abell clusters, while the BCS,
with its low-flux extension \citep{Ebel2}, has a flux limit of
$f_x > 2.8 \times 10^{-12}$ ergs sec$^{-1}$ cm$^{-2}$ and contains 304
clusters. Most of the listed ICM temperatures are estimated from the $L_x-T$
relation of \cite{WhiteF}, but a reasonable number of clusters have
measured temperatures [see Table 1 of \cite{WhiteF}]. The flux-limited
nature of these samples create an apparent $L_{x}-z$ relation,
as can be seen in Figure 1, and in order to avoid reproducing an
artificial redshift evolution we construct volume
limited subsamples (the delineated regions of Fig.1); one is a low-luminosity
($L_{x}>10^{44}$ ergs/sec) subsample, which spans a
limited redshift range ($z\lesssim 0.07$ and 0.09 for the XBACs and BCS,
receptively) and one high-luminosity subsample
($L_{x}> 8 \times 10^{44}$ ergs/sec), spanning a much larger
redshift range ($z\lesssim 0.19$ and 0.24 for the XBACs and BCS, respectively).

\section{Results \& Discussion}
Figure 2 shows the distribution of APM cluster ellipticities
as a function of redshift ($z<0.18$). The filled circles represent APM 
clusters that are also in the Abell/ACO catalogue. 
There is a definite trend of ellipticity with redshift
in the direction expected from 
an evolution of the dynamical status of 
clusters, supporting similar claims of \citet{MelC}.
Table 1 summarizes the quantitative correlation results for all tests.
The Pearson correlation coefficient
for the $\epsilon-z$ correlation is $r\simeq 0.2$ and
with a probability of being a chance correlation of
${\cal P}\simeq 10^{-8}$. This result is robust to changes of the
sample size by factors of two or three, 
depending on whether we use clusters with observed
redshifts or different cluster richness (see Table 1).

Tests were performed to investigate whether systematic effects 
in the shape-parameter
determination method could be responsible for the ellipticity-redshift
trend. Already in \cite{BPM} a detailed analysis of the performance of
the method as a function of sampling, distance and the presence of a
projected random background galaxy distribution 
[as predicted by the luminosity function of APM galaxies; \cite{Mad96}]
showed that such
effects will tend to overestimate by $\sim 0.1-0.15$ the ellipticity
of nearly spherical clusters (more so for distant clusters - but only by 5\%
more), they will underestimate the ellipticity of flat clusters, more
so at large-redshifts, while they will leave unaffected the clusters with
ellipticities around the mean of the distribution (ie., $\epsilon \sim
0.4$). These effects could in principle bias the ellipticity
redshift correlation, depending on the relative abundance of different
ellipticity clusters. 
We investigate this possibility by performing a Monte-Carlo 
simulation analysis in which we
assume the same underlying cluster ellipticity distribution at all
redshifts. Then using the individual APM cluster distance and richness 
characteristics we simulate, for each APM cluster, 
1000 mock clusters having ellipticities derived 
by randomly sampling
the corrected, for different biases, ellipticity distribution [see
Fig.6 of \citet{BPM}].
Mock clusters are constructed having King profiles
and the APM richness at the selected distance from the observer
as in \cite{BPM}. The estimated background is overlayed on the cluster image,
allowing however for clustering (we take the extreme case that all the 
background galaxies are clustered at the distance of the cluster). 
From the resulting population of $\sim$900000 cluster
ellipticities, which have the correct underlying APM redshift selection
function, we randomly select $N$-times 900 ellipticities 
and estimate their correlation with redshift 
($N=40000$). We find no significant
correlation (ie., $\langle r\rangle=0.08$ and $\langle{\cal
P}\rangle=0.08$), while only in 16 out of the 40000 cases do we find a similar 
$\epsilon-z$ correlation having the observed, or higher
significance level. This number 
drops to 3 if we use a random, instead of a clustered,
background in the mock cluster construction. Since we have used the
extreme case where all the background is assumed to be clustered at
the distance of each APM cluster, we should consider the derived
significance as an upper limit. Therefore the true significance of 
the observed $\epsilon-z$ correlation, taking
into account the possible redshift dependent systematic effects, is 
$${\cal P}_{\epsilon-z}\lesssim 4\times 10^{-4} \;\;,$$
verifying that our ellipticity determination method,
coupled with shot-noise and the projection of a clustered background
cannot create the observed $\epsilon-z$ trend.

Regarding the $kT -z$ relation, we have found that there is indeed a
correlation, especially apparent in the high-$L$ subsample, which
spans a larger range in $z$ (see Figure 3 and Table 1). 
The correlation coefficient for the combined XBAC-BCS 
high-$L$ subsample (were for the common
clusters we have used only the more accurate BCS values)
is $r\simeq 0.47$ with probability of chance correlation being
${\cal P}\simeq 3 \times 10^{-4}$. This correlation 
is also apparent in each individual cluster sample as
well, although with a slightly lower significance.
A similar but weaker $kT -z$ correlation is found in the low-$L$
sample.

Note, however, that many of the used temperatures are
based on the $L_x-kT$ relation of \citet{WhiteF} and thus this correlation
could be a manifestation of an underlying $L_x-z$ relation. 
In order to investigate this possibility 
we restrict our analysis only to those clusters with measured
temperatures, reducing our samples considerably (by more than 50\%),
but the correlation of $kT$ with $z$ persists (see Table 1). 
Especially for the
low-$L_x$ sample the correlation now is even stronger and more
significant. For the high-$L$ sample, where the sample
number has dropped by a factor of 3, there is
still an apparent correlation but the significance has dropped
considerably. Therefore, we confirm that cluster temperatures
do evolve with redshift in the recent past.

We return to the question of a possible recent $L_x$ evolution, which
if present it could extend, through the $L_x-kT$ relation, the
already existing $kT-z$ relation also to clusters with estimated
temperatures. Indeed, as seen in Figure 4, where we plot the $L_x$
versus $z$ (see also Table 1), we do find such a relation with
correlation coefficient $r\simeq 0.4$ and significance ${\cal P} \sim
0.003$ - 0.14 (depending on the subsample). Note, however, that if we
use all the clusters in our subsample (not only those with measured
temperatures) then we observe a correlation only for the high-$L_x$
sample ($r\simeq 0.2$) which is weak and of low significance (${\cal P} \sim
0.15$). It could be that clusters with measured temperatures have more
accurately measured luminosities.

\subsection{Conclusions}
We have presented evidence, based on
the optical APM and X-ray XBAC and BCS cluster samples, 
of a recent dynamical evolution of clusters,
a fact to be expected if in clusters the merging and anisotropic
accretion of matter along filaments have stopped and relaxation
processes are now in effect.
Gravitational relaxation will tend to isotropize the
clusters reducing their ellipticity, their ICM temperature (shock heated
during the violent merging phase of their formation) and their X-ray
luminosity. We have found indications of varying
strength and significance for all
three of these effects, supporting similar claims of \citet{MelC}, on
the basis of an ellipticity-redshift relation. These results appear to
be consistent with the expectations of a low-$\Omega_{m}$ Universe.

\acknowledgments

I would like to thank Adrian Melott for a discussion.

\begin{figure}
\plotone{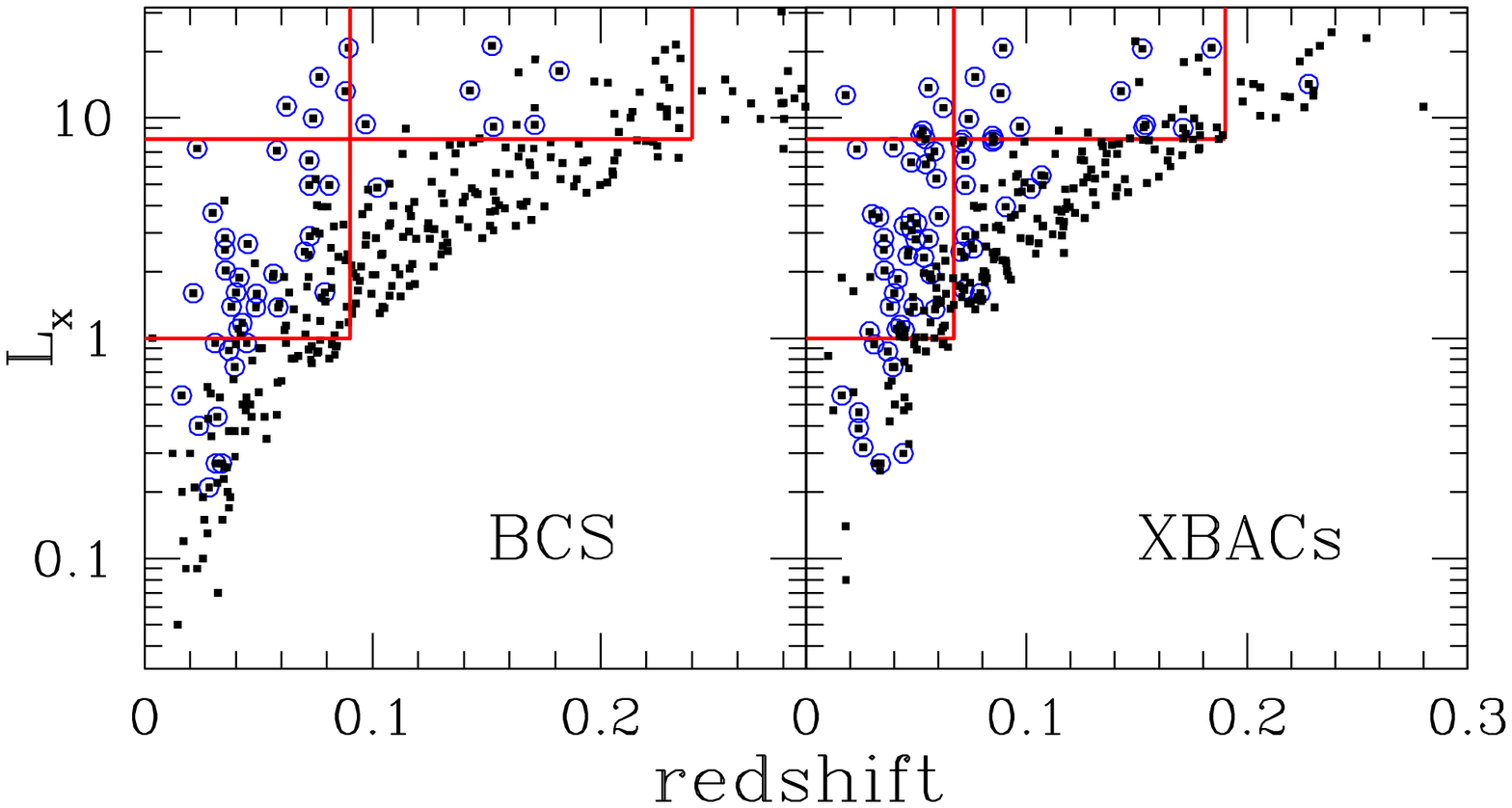}
\caption{The X-ray Luminosity-redshift plot for the BCS and XBAC
samples. The strong dependence of $L_x$ on $z$ due to the
flux limit is apparent. Open circles represent clusters with observed
temperatures from \cite{WhiteF} and the delineated regions represent
the volume-limited subsamples analysed.}
\end{figure} 


\begin{figure}
\plotone{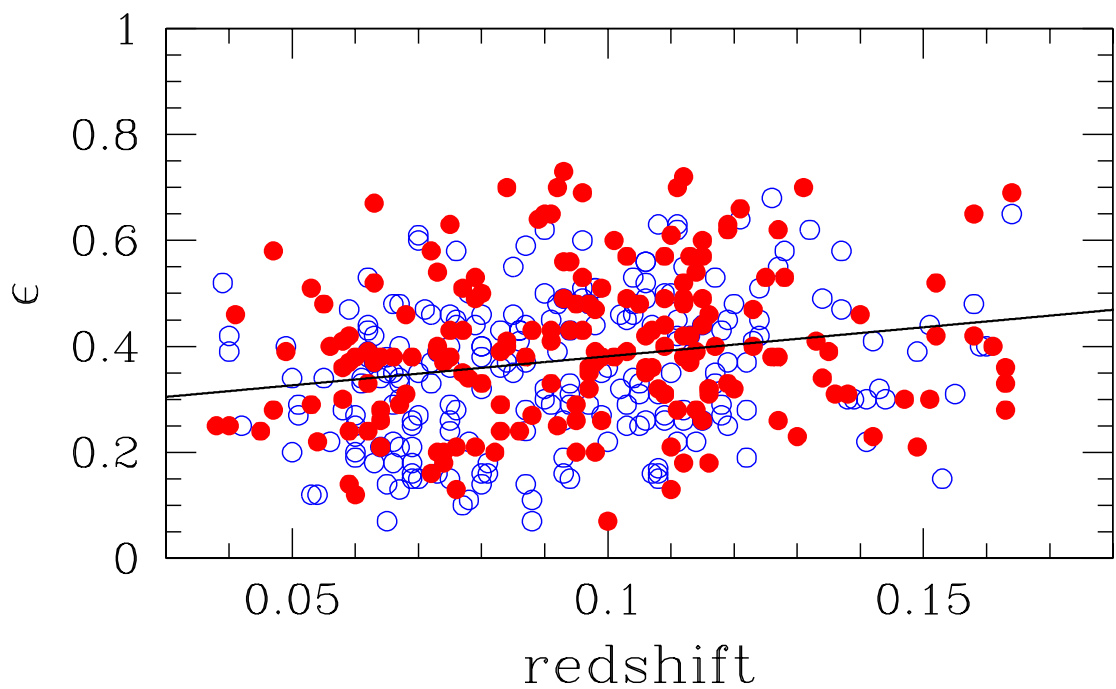}
\caption{The evolution with redshift of the APM cluster
ellipticity. For clarity only clusters with measured redshifts are
presented. Filled circles are those APM clusters that are also in the
Abell/ACO sample. The straight line represented the best least-square
fit to the data.}
\end{figure}

\begin{figure}
\plotone{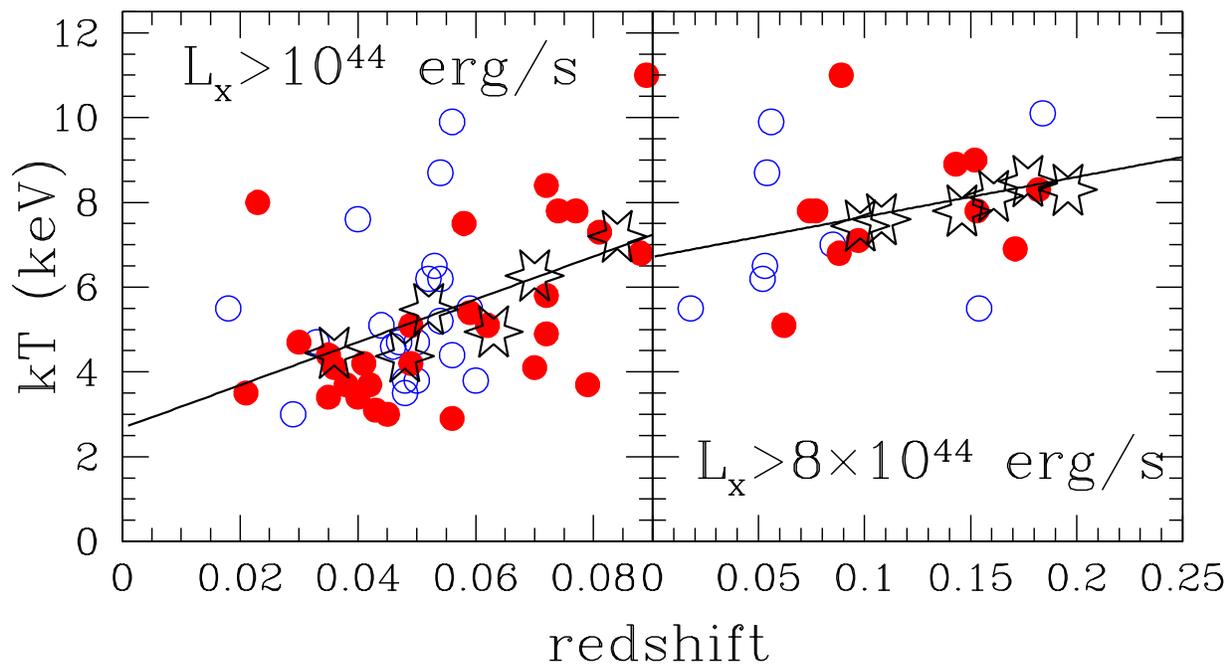}
\caption{The ICM temperature redshift correlation for the two
subsamples analysed. Filled and open points represent BCS and XBAC
clusters, respectively. The solid line is the best least-square fit to
the data, while the stars represent the mean ICM temperature in equal
volume shells (3 shells for each sample, plotted together). 
Note that we plot only those clusters with measured
temperatures from \citet{WhiteF} (those having errorbars in their
Table 1).}
\end{figure} 

\begin{figure}
\plotone{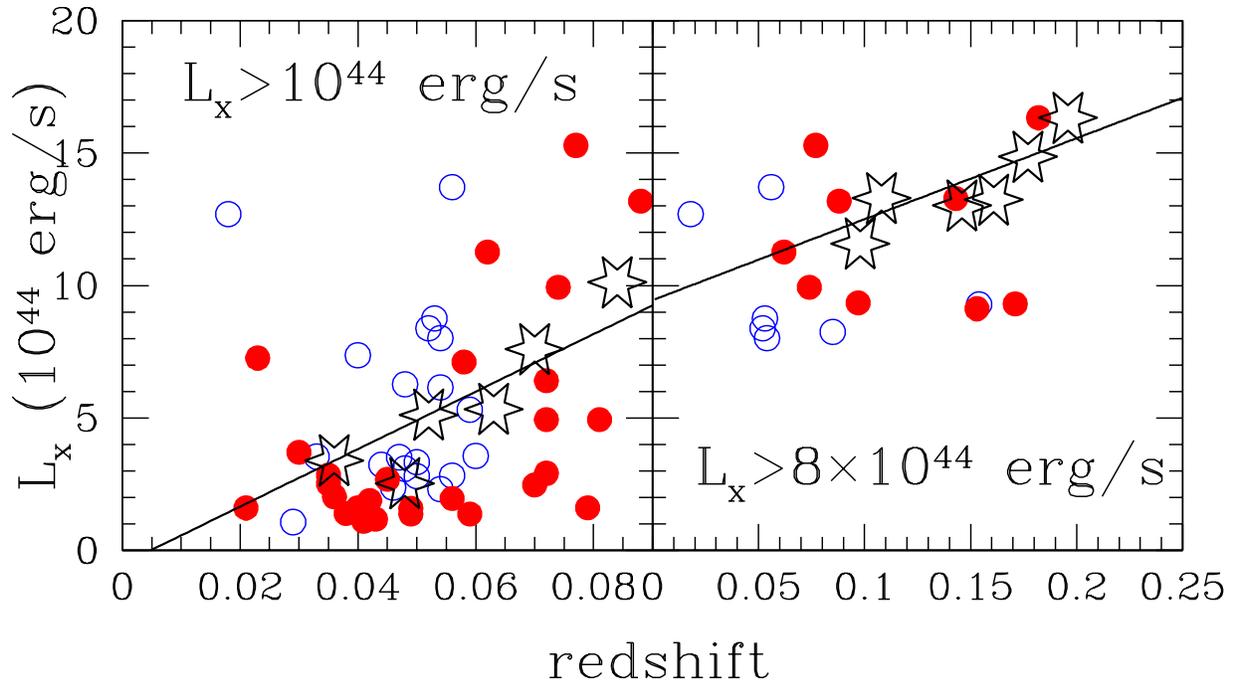}
\caption{The X-ray luminosity ($L_x$) redshift correlation for the two
subsamples analysed. Symbols are as in figure 3.} 
\end{figure} 

\clearpage

\begin{deluxetable}{lcccc}
\tabletypesize{\scriptsize}
\tablecaption{Correlation analysis results for the different
correlation pairs and samples.}
\tablewidth{0pt}
\tablehead{
\colhead{Cluster sample} & \colhead{$N$} & \colhead{correlation pair} & 
\colhead{$r$} & \colhead{${\cal P}$}
}
\startdata
APM ($z_{\lim} \sim 0.18$) & 899 & $\epsilon-z$ & 0.19 & $4.3\times 10^{-9}$ \\
APM ($z_{obs}$) & 404   & $\epsilon-z$ & 0.22 & $9.6\times 10^{-6}$ \\
APM ($>50$ counts) & 411 &$\epsilon-z$ & 0.21 & $2.8\times 10^{-5}$ \\
BCS+XBACs (high-$L_{x}$) & 55  &  $kT-z$ & 0.47 & $1.3\times 10^{-4}$ \\
BCS+XBACs (low-$L_{x}$) & 106  &  $kT-z$ & 0.16 &  $7.5\times 10^{-2}$ \\
BCS+XBACs (high-$L_{x}$) meas. & 19  &  $kT-z$ & 0.30 & 0.22 \\
BCS+XBACs (low-$L_{x}$) meas. & 48  &  $kT-z$ & 0.45 &  $1.5\times 10^{-3}$
\\
BCS+XBACs (high-$L_{x}$) meas. & 19  &  $L_x-z$ & 0.36 & 0.14 \\
BCS+XBACs (low-$L_{x}$) meas. & 48  &  $L_x-z$ & 0.42 &  $3\times 10^{-3}$
\\

\enddata
\end{deluxetable}

\begin{thebibliography}{}
\bibitem[Basilakos et al.(2000)]{BPM} Basilakos, S., Plionis, M.,
Maddox, S., 2000, \mnras, 316, 779
\bibitem[Beisbart, Valdarnini \& Buchert(2002)]{Bei}Beisbart, C., 
Valdarnini, R., Buchert, T., 2002, \aap, 379, 412
\bibitem[e.g., Buote(1998)]{Buote} Buote, D.A., 1998, \mnras, 293, 381 
\bibitem[Dalton et al.(1997)]{Dalton}Dalton G. B., Maddox S. J.,
Sutherland W. J., Efstathiou G., 1997,  \mnras, 289, 263 
\bibitem[Ebeling et al.(1996)]{Ebel} Ebeling, H., Voges, W.,
B\"ohringer, H., Edge, A.C., Huchra, J.P. and Briel, U.G., 1996,
\mnras, 281, 799
\bibitem[Ebeling et al.(1998)]{Ebel1} Ebeling, H., Edge, A.C., 
B\"ohringer, H., Allen, S.W., Crawford, C.S., Fabian, A.C., 
Voges, W., Huchra, J.P., 1998, \mnras, 301, 881
\bibitem[Ebeling et al.(2001)]{Ebel2} Ebeling, H., Edge, A.C., 
Allen, S.W., Crawford, C.S., Fabian, A.C., Huchra, J.P., 2001, \mnras, 318, 333
\bibitem[Evrard et al.(1993)]{Ev}Evrard A.E., Mohr J.J., Fabricant
D.G., Geller M.J., 1993, \apj, 419, L9
\bibitem[Kolokotronis et al.(2001)]{KBPG}  Kolokotronis, V.,
Basilakos, S., Plionis, M., Georgantopoulos, 2001, \mnras, 320, 49
\bibitem[Lacey \& Cole(1996)]{Lac} Lacey, C., Cole, S., 1996, \mnras, 262, 627
\bibitem[Lahav et al.(1991)]{Lah91}  Lahav, O., Rees, M.J., Lilje,
P.B., Primack, J., 1991, \mnras, 251, 128
\bibitem[Maddox et al.(1990)]{Mad90} Maddox S.J., Sutherland W.J.,
Efstathiou G., Loveday, J. 1990, \mnras, 243, 692
\bibitem[Maddox et al.(1996)]{Mad96} Maddox S.J., Efstathiou, G.,
Sutherland W.J., 1996, \mnras, 283, 1227
\bibitem[Melott et al.(2001)]{MelC} Melott, A.L., Chambers, S.W.,
Miller, C.J., 2001, \apj, 559, L75
\bibitem[Peebles(1980)]{Pee80}Peebles, 
P.J.E., 1980, {\em Physical Cosmology}, Princeton Univ. Press
\bibitem[Plionis et al.(1991)]{PBF91} Plionis, M, Barrow, J.D. \&
Frenk, C.S. 1991, \mnras, 249, 662
\bibitem[Plionis \& Basilakos(2002)]{PB02} Plionis, M. \& Basilakos,
S., 2002, \mnras, 329, L47
\bibitem[e.g., Quintana \& Melnick(1982)]{Quin} Quintana, H. \& Melnick,
J. 1982, \aj, 87, 972
\bibitem[Richstone et al.(1992)]{Ri}Richstone, D., Loeb, A., Turner,
E.L., 1992, \apj, 393, 477
\bibitem[e.g., Ritchie \& Thomas(2002)]{Ritchi}Ritchie, B.W., 
Thomas, P.A., 2002, \mnras, 329, 675
\bibitem[e.g., Rose et al.(2002)]{Rose} Rose, J.A., Gaba, A.E.,
Christiansen, W.A., Davis, D.S., Caldwell, N., Hunstead, R.W.,
Johnston-Hollitt, M., 2002, \aj, 123, 1216
\bibitem[Shandarin \& Klypin(1984)]{Shan}Shandarin, S.F. \& Klypin,
A., 1984, \sovast 28, 491
\bibitem[Suwa et al.(2002)]{Suwa} Suwa, T., Habe, A., Yoshikawa, K., 
Okamoto, T., 2002, \apj,  in press, {\em astro-ph/0108308}
\bibitem[e.g. West, Jones \& Forman(1995)]{West} West, M.J., Jones, C.,
\& Forman, W., 1995, \apj, 451, L5
\bibitem[White et al.(1997)]{WhiteF} White, D. A., Jones, C., Forman,
W., 1997, \mnras, 292, 419
\end{thebibliography}
\end{document}